\documentclass{article}
\usepackage{amsmath,amssymb}
\usepackage{url}
\usepackage{graphicx}

\def\GeV{{\rm\ GeV}}

\def\<{\langle}
\def\>{\rangle}
\def\be{\begin{equation}}
\def\ee{\end{equation}}
\def\bea{\begin{eqnarray}}
\def\eea{\end{eqnarray}}

\begin{document}
\title{Two photon exchange amplitude with $\pi N$ intermediate states:\\ spin-1/2 and spin-3/2 channels}
\author{Dmitry~Borisyuk, Alexander~Kobushkin\\[2mm]
\it Bogolyubov Institute for Theoretical Physics,\\
\it 14-B Metrologicheskaya street, Kiev 03680, Ukraine}
\date{}
\maketitle

\begin{abstract}
 We calculate two-photon exchange (TPE) amplitudes for the elastic electron-proton scattering,
 and take into account intermediate hadronic states containing $\pi N$ system with total angular momentum 1/2 or 3/2, which includes 8 different channels.
 This is the improvement of our previous calculation, where only the $\pi N$ states with quantum numbers of $\Delta$ resonance were included.
 The results show good consistency with recent experimental data.
 At high $Q^2$, newly-calculated contributions affect the correction to the measured proton form factor ratio $\mu G_E/G_M$.
 The total correction becomes somewhat smaller compared to our previous work, but is still significant and grows approximately linearly with $Q^2$.
 Comparing contributions of different channels, we found that larger contributions come
 from the channels with quantum numbers of lightest resonances.
\end{abstract}

\section{Introduction}

In our previous work \cite{ourP33}, we presented the calculation of the two-photon exchange (TPE) amplitude
for the elastic $ep$ scattering with the $\pi N$ intermediate states with quantum numbers of $\Delta$ resonance ($P_{33}$ channel).
In the present paper, we report an extension of that calculation, which includes all $\pi N$ states
with spin-parities $J^P = 1/2^+$, $1/2^-$, $3/2^+$, $3/2^-$ and compare the results to recent experimental data.

The only paper that has a similarity to the present work is Ref.~\cite{BMTres},
where TPE contributions of spin-1/2 and spin-3/2 resonances were studied.
However, in Ref.~\cite{BMTres} the resonances were assumed to be infinitely narrow,
and only one observable, the unpolarized cross-section, was considered.
In the present work we naturally take into account finite resonance width, realistic resonance shape and form factors, as well as non-resonant background,
and study different observables.

For the details of our method, see Ref.~\cite{ourP33}.
Here we will briefly describe differences to the previously published material.

\section{Some background}

We follow the strategy described in Ref.~\cite{ourP33} to calculate the contributions of the $\pi N$ intermediate states.
The $\pi N$ channel with given (iso)spin and parity can be viewed as a continuum of resonances $R_W$ with those quantum numbers and varying mass $W$.
The TPE contribution of the fictitious resonance $R_W$ can be calculated as usually
and the full contribution of the corresponding channel is obtained after the integration over $W$. 
States with spin-parity $J^P = 1/2^\pm$, $3/2^\pm$ and isospin $I = 1/2$, $3/2$ are included,
which comprise the following eight channels: $S_{11}$, $S_{31}$, $P_{11}$, $P_{31}$, $P_{13}$, $P_{33}$, $D_{13}$, $D_{33}$.

The amplitude for the electromagnetic transition between some particle $R_W$ with mass $W$ and the proton with mass $M$,
\be
R_W(p) \to \gamma^*(q) + p(p-q)
\ee
(where the momenta are shown in parentheses) is, for $J = 3/2$
\be
\<p|J_\mu|R_W\> = \frac{1}{4M^2 \sqrt{MW}} \; (g_{\mu\alpha}q_\nu - g_{\nu\alpha}q_\mu) \;
  \bar U \left[ (\hat p \gamma^\nu - p^\nu) F_1 - p^\nu F_2 + q^\nu F_3 \right] \gamma_5^{\frac{1+P}{2}} V^\alpha
\ee
and for $J=1/2$
\be
\<p|J_\mu|R_W\> = \frac{1}{2\sqrt{MW}} \;
  \bar U \left[ \left( \gamma_\mu - \frac{q_\mu \hat q}{q^2} \right) F_1 + (\gamma_\mu \hat q - q_\mu) \frac{F_2}{2M} \right] \gamma_5^{\frac{1-P}{2}} V,
\ee
where $U$ and $V$ are spinors of the proton and the particle $R_W$, respectively, and $P=\pm 1$ is $R_W$'s parity.
The transition form factors $F_i$ depend on $q^2$ and $W$. They are complex, though the overall phase is irrelevant.
The relations of the transition form factors $F_i$ with the helicity electroproduction amplitudes $A_H$ ($A_{1/2}$, $A_{3/2}$ and $S_{1/2}$) are, for $J = 3/2$
\bea
K F_1 &=& [(PW-M)^2-q^2](PA_{3/2}+\sqrt{3}A_{1/2}) \nonumber \\
K F_2 &=& [W^2-M^2+q^2](PA_{3/2}-\sqrt{3}A_{1/2}) + 2 q^2 PW \sqrt{6} \frac{S_{1/2}}{|\vec q|} \nonumber \\
K F_3 &=& 2W^2(PA_{3/2}-\sqrt{3}A_{1/2}) + [W^2-M^2+q^2] PW \sqrt{6} \frac{S_{1/2}}{|\vec q|}
\eea
and for $J=1/2$
\bea
K F_1 &=& -\frac{q^2}{2M^2} [(PW-M)^2-q^2] \left[ \frac{A_{1/2}}{\sqrt{2}} + (PW+M)\frac{S_{1/2}}{|\vec q|} \right] \nonumber \\
K F_2 &=& \frac{1}{M} [(PW-M)^2-q^2] \left[ (PW+M)\frac{A_{1/2}}{\sqrt{2}} + q^2 \frac{S_{1/2}}{|\vec q|} \right] 
\eea
where $K$ is the same as in Eq.(7) of Ref.~\cite{ourP33}.
The above equations supersede Eq.(6) of Ref.~\cite{ourP33}.

The numerical values of the amplitudes $A_H$ were taken from the MAID model \cite{MAID}.

We need to make a remark on the isospin structure of the amplitudes.
For our purposes we need the amplitudes for the transition to states of definite isospin ($I=1/2$ or $I=3/2$), they are 
\bea
A(\gamma^*p \to \pi N|_{I=3/2}) &=& \frac{\sqrt{2}}{\sqrt{3}} A_{p\pi^0} - \frac{1}{\sqrt{3}} A_{n\pi^+} = \frac{\sqrt{2}}{\sqrt{3}} A^{(3/2)} \\
A(\gamma^*p \to \pi N|_{I=1/2}) &=& -\frac{1}{\sqrt{3}} A_{p\pi^0} - \frac{\sqrt{2}}{\sqrt{3}} A_{n\pi^+} = -\sqrt{3} A_p^{(1/2)}
\eea
where the amplitudes in the r.h.s. are those defined in Ref.~\cite{MAID}.

The calculation is then done in the same way as in Ref.~\cite{ourP33},
with the updated version of {\tt TPEcalc} program, which now supports intermediate states with spin-parities $1/2^\pm$, $3/2^\pm$ \cite{TPEcalc}.

\section{Results}

\subsection{$R_\pm$ experiments} 
There were two recent experiments to search for TPE effects: \cite{VEPP,CLAS}.
In both experiments, the positron-to-electron cross-section ratio $R_\pm = \sigma(e^+p)/\sigma(e^-p)$ was measured.
This ratio is strictly equal to 1 in one photon exchange approximation,
thus allowing direct observation of TPE contribution to the unpolarized cross-section.

Results of the VEPP-3 experiment \cite{VEPP} show clear deviation of $R_\pm$ from unity.
The assumption of ``no TPE" yields bad $\chi^2$, and is thus inconsistent with data.
The authors analyzed several different models of TPE effects, and had found
that their experimental results agree best with Refs.~\cite{BMT,ourDisp} and \cite{Bernauer}
(note however that Ref.~\cite{Bernauer} estimates TPE effect from an analysis of experimental data, not from a theoretical calculation).

We have calculated $R_\pm$ for the kinematical conditions of Ref.~\cite{VEPP}.
The calculation was done using several different ``flavours" of our model:
\begin{itemize}
\item with the elastic intermediate state only \cite{ourDisp}
\item with elastic + $\Delta$ resonance with zero width \cite{ourDelta}
\item with elastic + $\pi N$ ($P_{33}$ channel), which includes $\Delta$ with realistic width and background \cite{ourP33}
\item with elastic + eight $\pi N$ channels as in the present work
\end{itemize}

\begin{figure}[t]
 \includegraphics[width=0.5\textwidth]{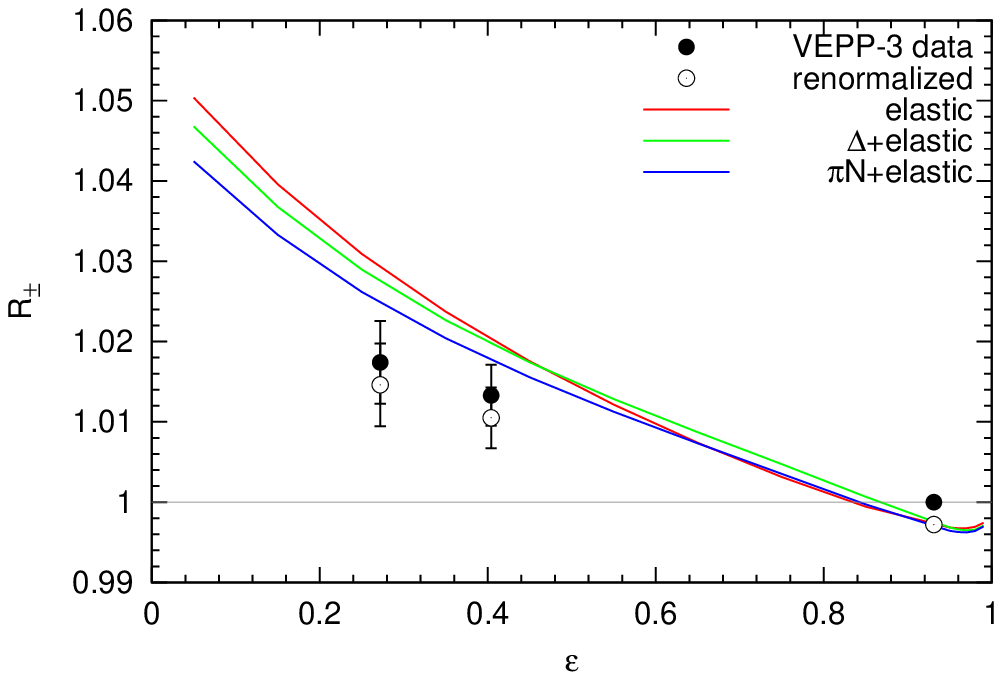}
 \includegraphics[width=0.5\textwidth]{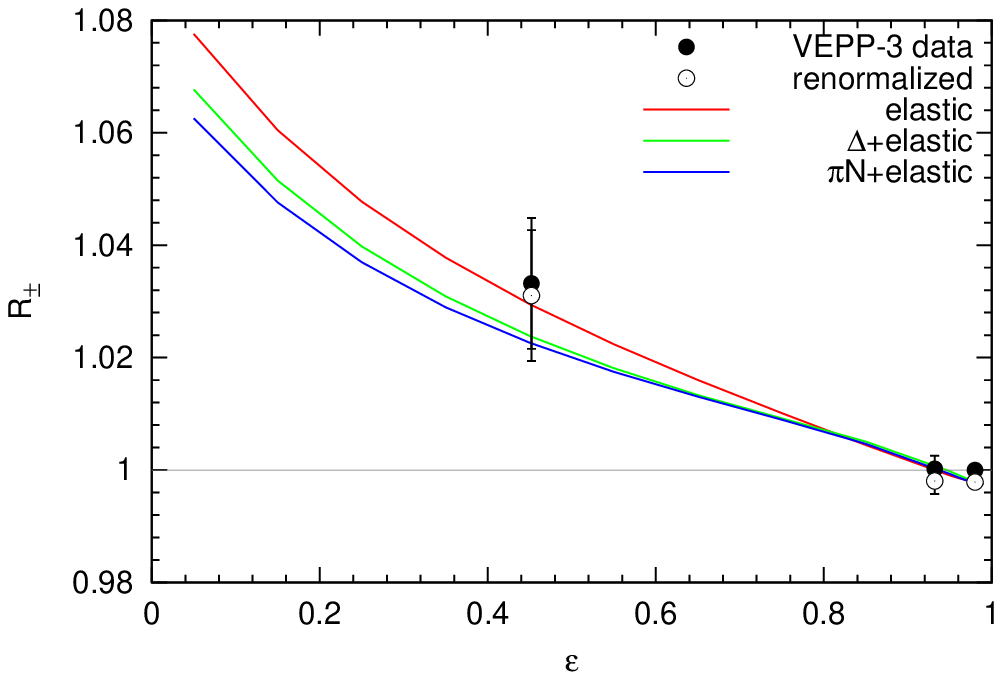}
 \caption{Comparison of VEPP-3 results \cite{VEPP} and our calculations. $E = 1.0\GeV$ (left), $E=1.6\GeV$ (right).}\label{Fig:VEPP}
\end{figure}

There are two approaches to comparison of theory and experiment, used in Ref.~\cite{VEPP}.
As TPE corrections should vanish at $\epsilon \to 1$, the experimental points with highest $\epsilon$
were used to determine normalization, i.e. $R_\pm$ was assumed to be 1 there.
Then the theoretically calculated $R_\pm$ is compared to thus normalized data
(first row of Table~\ref{Tab:VEPP} and black circles in Fig.~\ref{Fig:VEPP}).
The alternative, ``renormalization", approach is to shift data so that $R_\pm$ become equal to 
the theoretical prediction at that points (second row of Table~\ref{Tab:VEPP} and white circles in Fig.~\ref{Fig:VEPP}).
We see that, with both approaches, the agreement gets gradually better as we include more intermediate states.
The best $\chi^2$ is achieved with the first approach and TPE from the present work --- we have $\chi^2$/d.o.f. = 1.06.

\begin{figure}
 \includegraphics[width=0.5\textwidth]{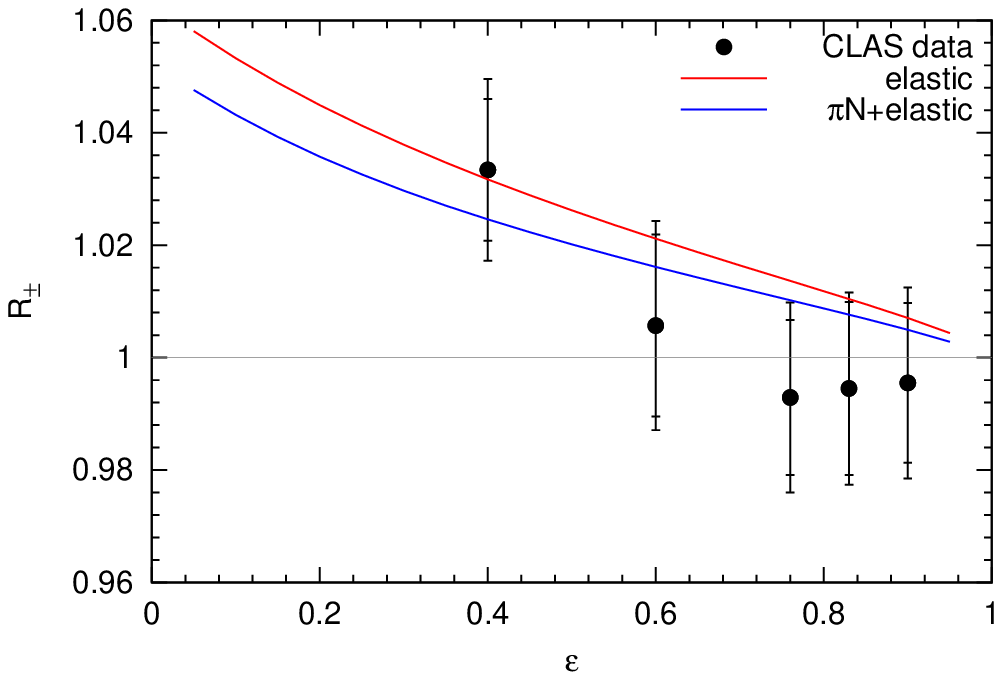}
 \includegraphics[width=0.5\textwidth]{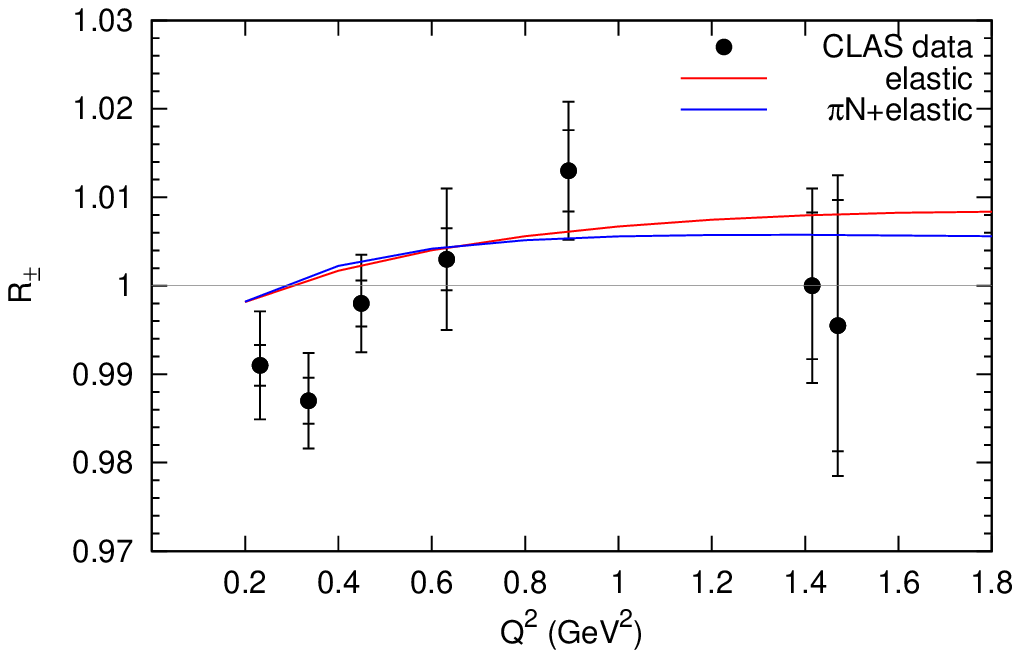}
 \caption{Comparison of CLAS results \cite{CLAS} and our calculations. $Q^2 = 1.45\GeV^2$ (left), $\epsilon = 0.88$ (right).}\label{Fig:CLAS}
\end{figure}

As for the CLAS experiment \cite{CLAS}, the results for $R_\pm$ quoted there
are not so far from 1 within errors, i.e. ``no TPE" hypothesis is not clearly rejected
(the authors mention $2.5\sigma$ preference to TPE over ``no TPE").
Nevertheless, we see similar gradual improvement of $\chi^2$ with the improvement of TPE model (Table~\ref{Tab:VEPP}).
The corresponding curves are plotted in Fig.~\ref{Fig:CLAS}.
\begin{table}[b] \center
  \begin{tabular}{|c|cccc|c|}
\hline
     Data & no TPE & elastic \cite{ourDisp} & narrow $\Delta$ \cite{ourDelta} & full $P_{33}$ \cite{ourP33} & this work \\
\hline
     VEPP           & 7.97 & 2.19  &  1.86 & 1.68 & 1.06 \\
     VEPP, renorm.  & 7.97 & 3.87  &  3.37 & 3.18 & 2.44 \\
     CLAS           & 1.43 & 1.24  &  1.29 & 1.22 & 1.21 \\
\hline
  \end{tabular}
\caption{$\chi^2$/d.o.f. for the comparison of VEPP-3 and CLAS experiments with different TPE calculations.}\label{Tab:VEPP}
\end{table}

\subsection{GEp$2\gamma$ experiment}

In GEp$2\gamma$ experiment \cite{GEP} the $\epsilon$ dependence of measured proton form factor ratio $R$ at $Q^2 = 2.5\GeV^2$ was studied.
In one photon exchange approximation this quantity is equal to $\mu G_E/G_M$, and thus independent of $\epsilon$,
so a variation of $R$ with $\epsilon$ is (a sign of) TPE effect. The authors conclude that experimental data show no evidence for such a variation.

A comparison of our calculations with the data is shown in Fig.~\ref{Fig:GEP},
again in two approaches. The dashed lines correspond to the case where the value of $R$ at $\epsilon=1$,
which we need to add to the calculated TPE correction, is taken as average of experimental data (dashed lines).
In the second case (solid lines) this value is free parameter, determined by fitting.
The agreement of data and theory is evidently better in this case; we see that our results do not conflict with experimental data. 

\subsection{High $Q^2$}

The contribution of the newly-calculated intermediate states to the {\it unpolarized cross-section}
is rather small at high $Q^2$; the elastic contribution dominates in this case.

\begin{figure}
\parbox[t]{0.49\textwidth}{
  \includegraphics[width=0.5\textwidth]{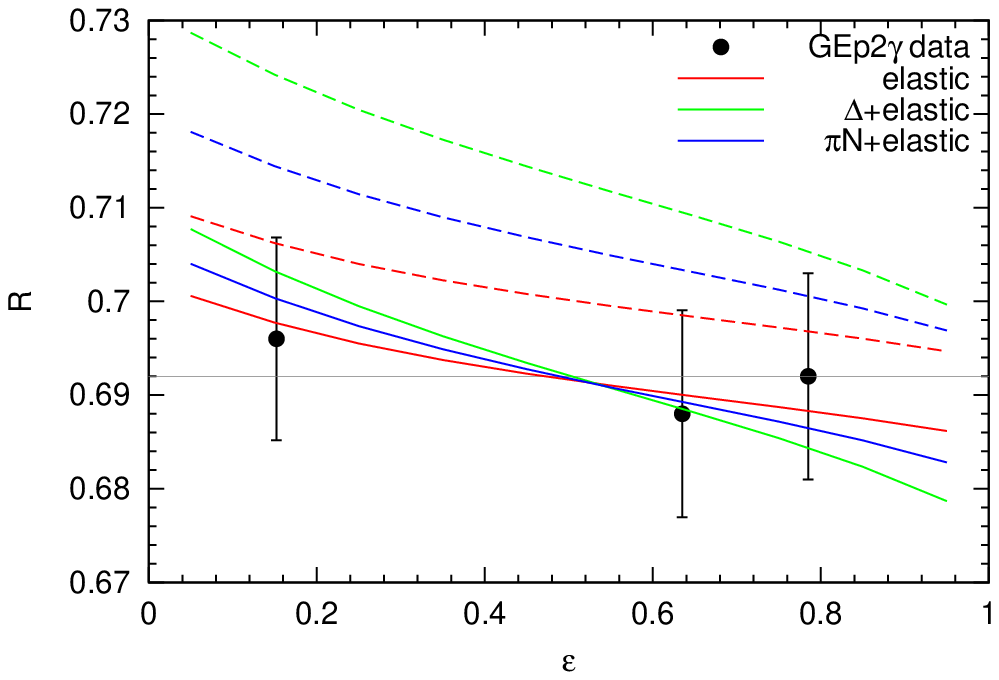}
  \caption{Comparison with GEp$2\gamma$ experiment \cite{GEP}. Dashed lines: $R|_{\epsilon=1}$ fixed, solid lines: $R|_{\epsilon=1}$ fitted.}\label{Fig:GEP}
}\hfill
\parbox[t]{0.49\textwidth}{
  \includegraphics[width=0.5\textwidth]{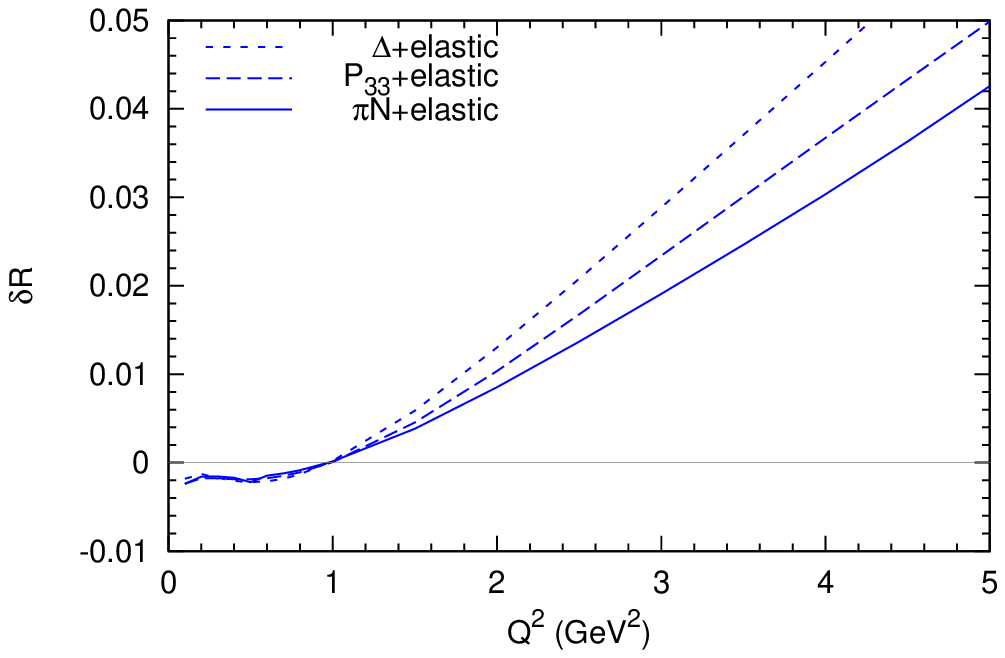}
  \caption{Full TPE correction to measured proton form factor ratio at high $Q^2$, fixed $\epsilon = 0.5$.}\label{Fig:high}
}
\end{figure}

It was found in Refs.~\cite{ourDelta,ourP33}, that the TPE contribution to the measured {\it proton form factor ratio} $R$,
coming from either $\Delta$ resonance or full $\pi N (P_{33})$ channel, is increasing with $Q^2$ and far exceeds the elastic contribution.
Adding the contributions from the channels considered in the present work, we see that the total effect,
though becomes somewhat smaller, is still sizeable, and the overall trend persists:
$\delta R$ grows approximately linearly with $Q^2$, Fig.~\ref{Fig:high}.

\subsection{Relative size of the contributions of different channels}

\begin{figure}
 \includegraphics[width=0.5\textwidth]{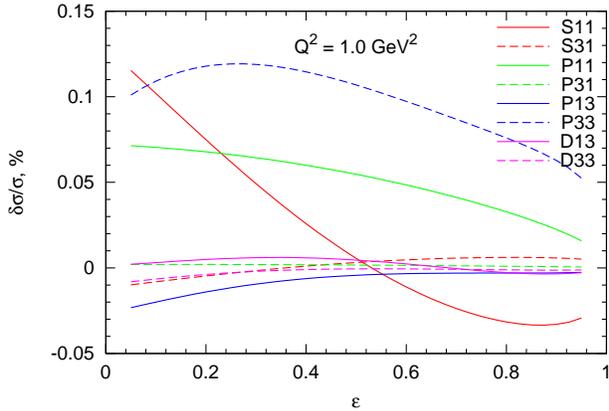}
 \includegraphics[width=0.5\textwidth]{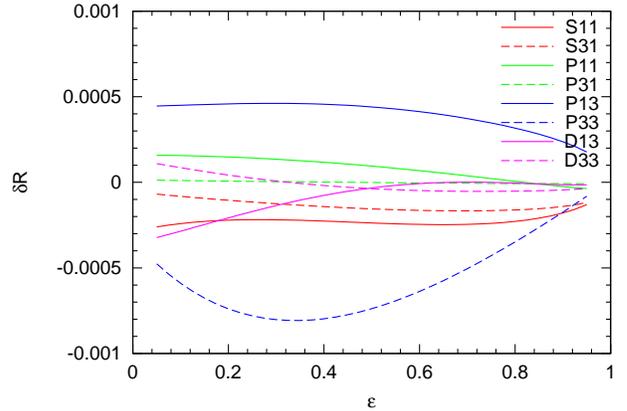}
 \includegraphics[width=0.5\textwidth]{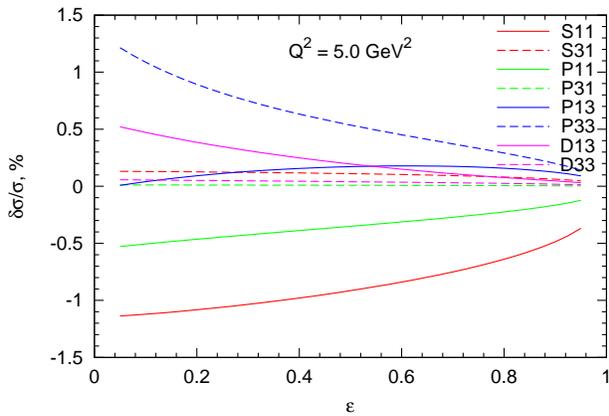}
 \includegraphics[width=0.5\textwidth]{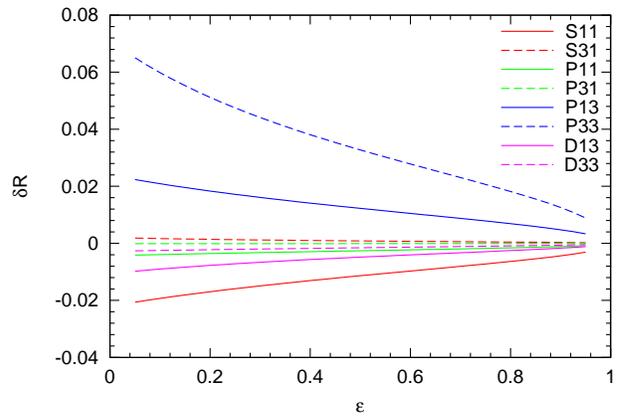}
 \caption{TPE contributions of different channels compared, fixed $Q^2$ as indicated on the plots.}\label{Fig:cmpQ2}
\end{figure}

\begin{figure}
 \includegraphics[width=0.5\textwidth]{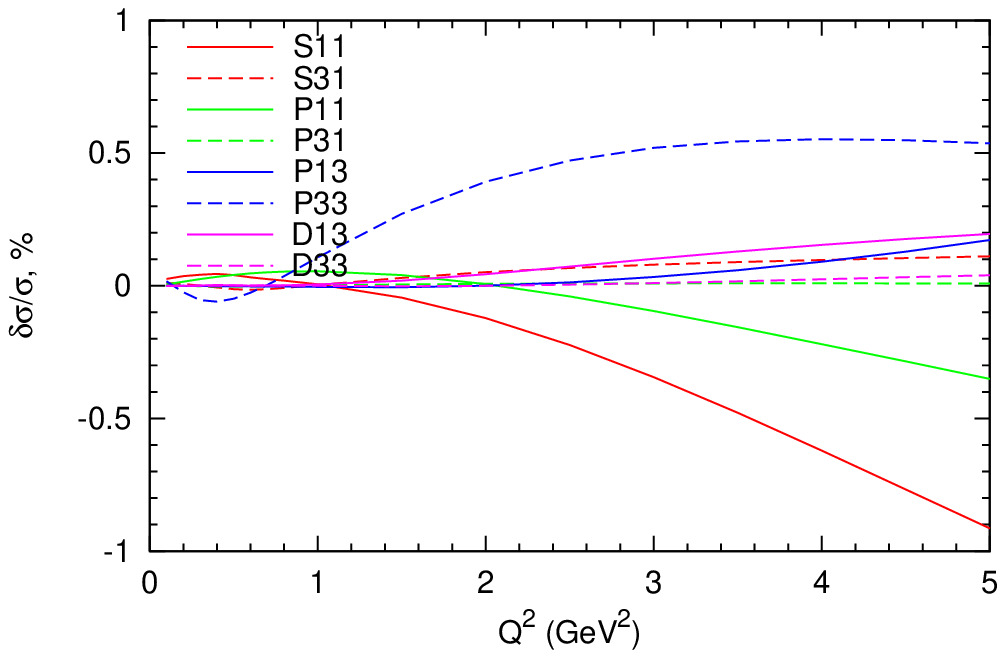}
 \includegraphics[width=0.5\textwidth]{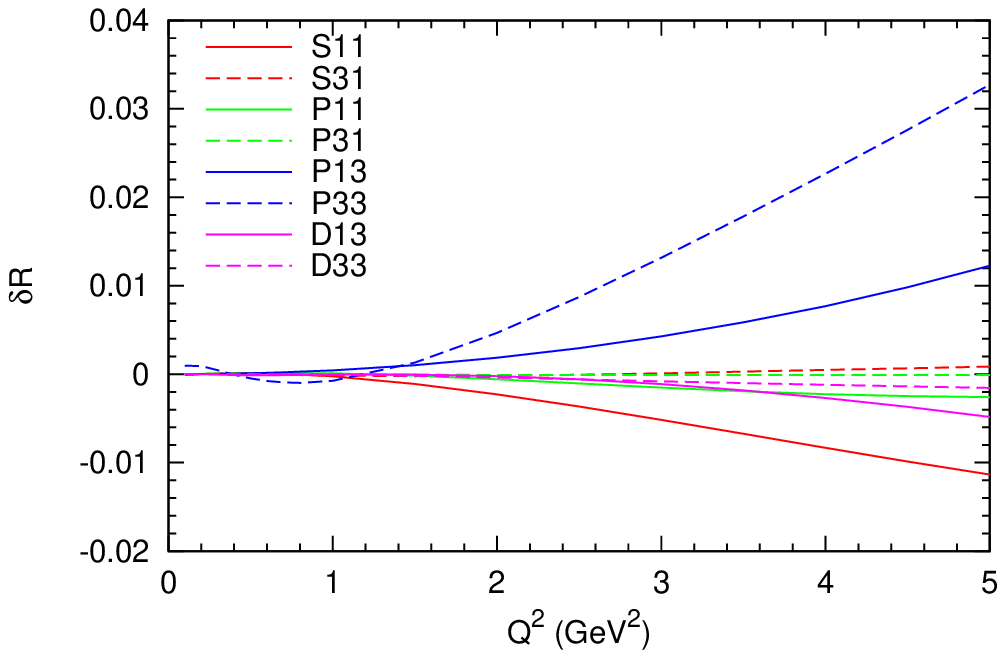}
 \caption{TPE contributions of different channels compared, fixed $\epsilon=0.5$.}\label{Fig:cmpEPS}
\end{figure}

As we have calculated the contributions of 8 distinct channels, it is interesting to compare them with each other.
In Figs.~\ref{Fig:cmpEPS},\ref{Fig:cmpQ2} we plot corresponding contributions to the unpolarized cross-section, $\delta\sigma/\sigma$,
and measured form factor ratio, $\delta R$, versus $Q^2$ and $\epsilon$.

We see that the dominant contribution among all $\pi N$ channels always belongs to $P_{33}$, related to $\Delta(1232)$ resonance.
Other significant contributions come from $S_{11}$, $P_{11}$ (only in $\delta\sigma/\sigma$),
$P_{13}$ (only in $\delta R$), and in some cases $D_{13}$, channels.
These are the channels where lightest resonances are found ---
$S_{11}(1535)$, $D_{13}(1520)$, and Roper resonance $P_{11}(1440)$.
The findings of Ref.~\cite{BMTres} were somewhat different: $D_{13}$ yielded next largest contribution after $P_{33}$,
and the effect of other resonances was almost negligible.
We think this is due to the limitations of the approach of Ref.~\cite{BMTres}:
the Roper resonance has a large width, and $S_{11}$ channel has significant non-resonant contribution near threshold,
but both are absent in the approximation of Ref.~\cite{BMTres}.

In qualitative agreement with Ref.~\cite{BMTres} (this was also suggested in Ref.~\cite{ourTNSA}),
we see that the contributions of different channels tend to cancel each other.

Nevertheless, some doubt remains. The contributions calculated so far
can be viewed as the first terms of the (infinite) expansion of the total $\pi N$ contribution
in intermediate state's spin $J$. Whether the series is convergent, is not fully clear.
Some light may be shed onto this question by calculating contributions
of the intermediate states with higher spins ($J \ge 5/2$), which is currently underway.

\section{Conclusions}

We have calculated TPE amplitudes for the elastic electron-proton scattering,
taking into account, besides the elastic intermediate state, intermediate states containing $\pi N$ system with total angular momentum 1/2 or 3/2.
This corresponds to 8 channels ($S_{11}$, $S_{31}$, $P_{11}$, $P_{31}$, $P_{13}$, $P_{33}$, $D_{13}$, $D_{33}$).
The results agree well with the recent experimental data \cite{VEPP,CLAS}.
The agreement is better than with the inclusion of the elastic intermediate state only \cite{ourDisp},
and better than with elastic + single $P_{33}$ channel \cite{ourP33},
i.e. adding new intermediate states improves the agreement with the experiments.

At high $Q^2$, newly-calculated contributions affect the correction to the measured proton form factor ratio $R$.
The total correction, $\delta R$, is 
smaller compared to Ref.~\cite{ourP33}, but still grows approximately linearly with $Q^2$.

Comparing contributions of different channels, we see that larger ones come
from the channels with quantum numbers of lightest resonances.
It would be interesting to check the contributions of channels with higher spins ($\ge 5/2$),
since they also contain prominent resonances, e.g. $F_{15}(1680)$.
Such work is currently in progress.

\end{document}